\documentclass[journal]{IEEEtran}

\usepackage{graphicx}
\usepackage{epsfig}
\usepackage{multirow}
\usepackage{times}
\usepackage{amsmath}
\usepackage{latexsym}
\usepackage{flushend}
\usepackage[noadjust]{cite}
\usepackage{subfigure}
\usepackage{booktabs}
\usepackage{array}
\usepackage{url}
\usepackage{amssymb}
\usepackage{float}

\ifCLASSINFOpdf
\else
\fi
\begin{document}
\title{Efficient VVC Intra Prediction Based on Deep Feature Fusion and Probability Estimation}
%
%
%

\author{Tiesong~Zhao,~\IEEEmembership{Senior Member,~IEEE,}~Yuhang~Huang,~Weize~Feng,~Yiwen~Xu,~\IEEEmembership{Member,~IEEE} and Sam Kwong,~\IEEEmembership{Fellow,~IEEE} 
\thanks{
This work is supported by the National Natural Science Foundation of China under Grant 62171134. (*Corresponding author: Yiwen Xu, xu\_yiwen@fzu.edu.cn.)

T. Zhao is with the Fujian Key Lab for Intelligent Processing and Wireless Transmission of Media Information, College of Physics and Information Engineering, Fuzhou University, Fuzhou 350108, China, and also with the Peng Cheng Laboratory, Shenzhen 518055, China (e-mail:
t.zhao@fzu.edu.cn)

Y. Huang, W. Feng and Y. Xu are with the Fujian Key Lab for Intelligent Processing and Wireless Transmission of Media Information, College of Physics and Information Engineering, Fuzhou University, Fuzhou 350108, China (e-mails: {n191127013, 201127019, xu\_yiwen}@fzu.edu.cn).

S. Kwong is with the Department of Computer Science, City University of Hong Kong, Hong Kong SAR. (e-mail: cssamk@cityu.edu.hk).
}}%
\maketitle

\begin{abstract}
The ever-growing multimedia traffic has underscored the importance of effective multimedia codecs. Among them, the up-to-date lossy video coding standard, Versatile Video Coding (VVC), has been attracting attentions of video coding community. However, the gain of VVC is achieved at the cost of significant encoding complexity, which brings the need to realize fast encoder with comparable Rate Distortion (RD) performance. In this paper, we propose to optimize the VVC complexity at intra-frame prediction, with a two-stage framework of deep feature fusion and probability estimation. At the first stage, we employ the deep convolutional network to extract the spatial-temporal neighboring coding features. Then we fuse all reference features obtained by different convolutional kernels to determine an optimal intra coding depth. At the second stage, we employ a probability-based model and the spatial-temporal coherence to select the candidate partition modes within the optimal coding depth. Finally, these selected depths and partitions are executed whilst unnecessary computations are excluded. Experimental results on standard database demonstrate the superiority of proposed method, especially for High Definition (HD) and Ultra-HD (UHD) video sequences.
\end{abstract}

\IEEEpeerreviewmaketitle

\section{Introduction}

\IEEEPARstart{V}{ideo} transmission has dominated the bandwidth of consumer internet \cite{001}. In particular, the user requirements on high fidelity pictures have promoted a booming of High Definition (HD) and Ultra HD (UHD) videos ({\it e.g.} 2K/4K/8K videos). To deliver these videos under limited bandwidth, the Joint Video Experts Team (JVET) developed a new video coding standard named as Versatile Video Coding (VVC), which has significantly improved the compression efficiency compared with its ancestors \cite{002}.

VVC inherits the block-based hybrid coding structure from earlier standards. In VVC, the input video frame is split into blocks called Coding Tree Units (CTUs). CTU consists of Coding Units (CUs) at different levels that share the same prediction style ({\it i.e.} intra or inter). The CU partition process is implemented by computing and comparing Rate-Distortion (RD) costs of all partitions, which is an extremely time-consuming task. Thus, the optimization on efficient CU partitioning is always a crucial problem in video coding.

Until now, there have been enormous contributions on efficient CU partitioning that were implemented on popular video codecs such as H.264/AVC and H.265/HEVC.  However, these methods cannot be directly transferred to the newly-developed VVC codec due to the complicated changes in coding structures. There is still lack of low complexity coding algorithms suitable for the latest version of VVC. Among the existing VVC algorithms, the low complexity intra prediction attracts less attention. To address this issue, we propose a two-stage framework in this paper.

In the light of complicated infrastructure of VVC, especially for its nested and hierarchical CTU structure, we extract spatial-temporal adjacent coding features that benefit the prediction of best coding depth and partition in consequent CUs. The state-of-the-art Convolutional Neural Network (CNN) and probability model facilitate this prediction process with high accuracy and low complexity. With these advantages, we propose a two-stage prediction framework for intra depth and partition mode of VVC.  The main contributions are summarized as follows.
\begin{enumerate}
	\item We propose an intra Depth prediction model with Deep Feature Fusion (D-DFF). Based on the spatial-temporal coherence of videos, this model fuses the spatial-temporal reference features at different scales and predicts the optimal depth of video with a deep CNN. 
	\item We propose an intra Partition mode prediction model with Probability Estimation (P-PBE). This model initializes the candidate partition modes and skips all unnecessary partitions for computational complexity reduction.
	\item The proposed two-stage framework is implemented in VVC and achieves promising complexity reduction and negligible RD loss compared with the state-of-the-art algorithms.
\end{enumerate}

The rest of this paper is organized as follows. Related works are provided in Section \uppercase\expandafter{\romannumeral2}. Section \uppercase\expandafter{\romannumeral3} presents the proposed two-stage framework with D-DFF and P-PBE. In Section \uppercase\expandafter{\romannumeral4}, we conduct the experiments to examine the effectiveness of our method. Finally, the conclusions of our works are given in Section \uppercase\expandafter{\romannumeral5}.

\section{Related Works}

\begin{figure*}[t!]
\centering
\subfigure[\label{Fig1a}]{
\begin{minipage}[t]{0.33\linewidth}
\centering
\includegraphics[width=4.5cm]{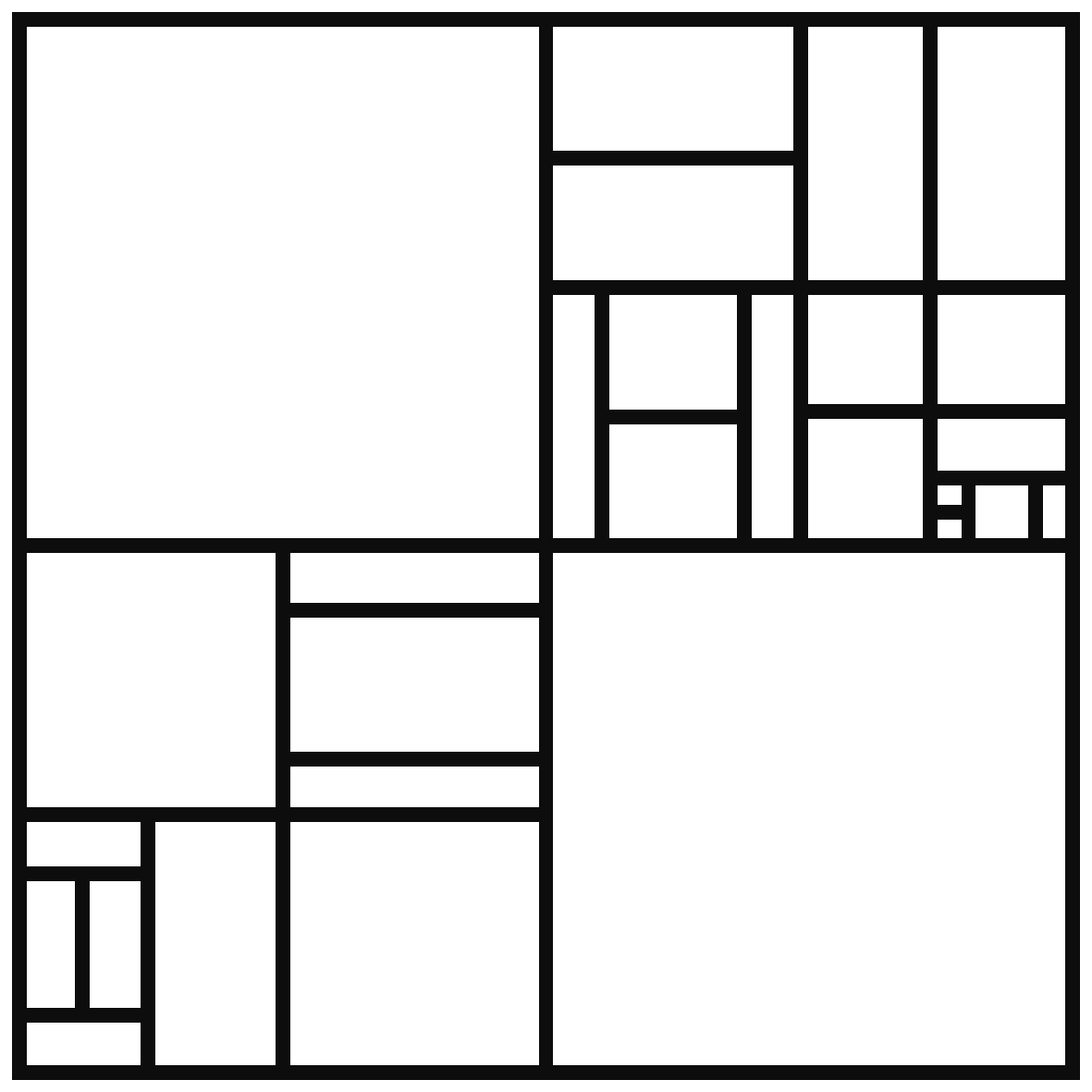}
\end{minipage}%
}%
\subfigure[\label{Fig1b}]{
\begin{minipage}[t]{0.33\linewidth}
\centering
\includegraphics[width=6.0cm]{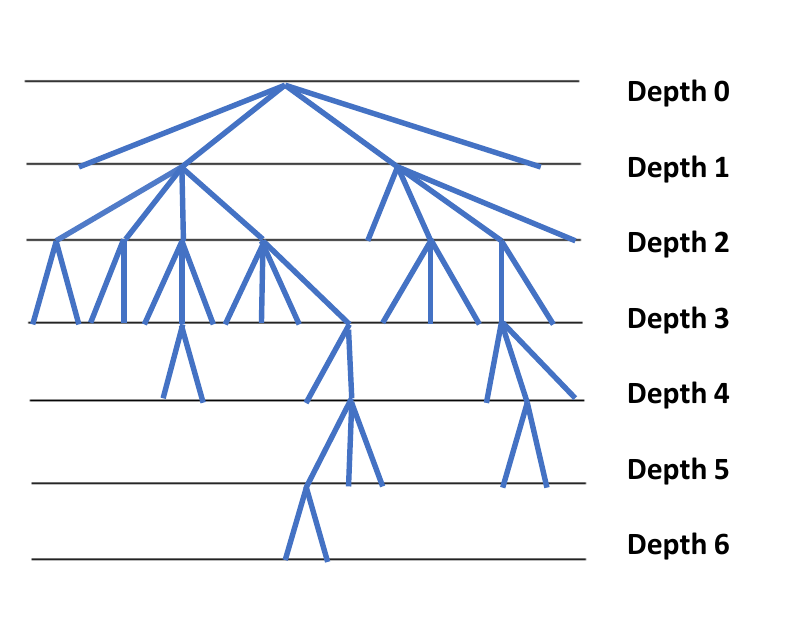}
\end{minipage}%
}%
\subfigure[\label{Fig1c}]{
\begin{minipage}[t]{0.33\linewidth}
\centering
\includegraphics[width=5.0cm]{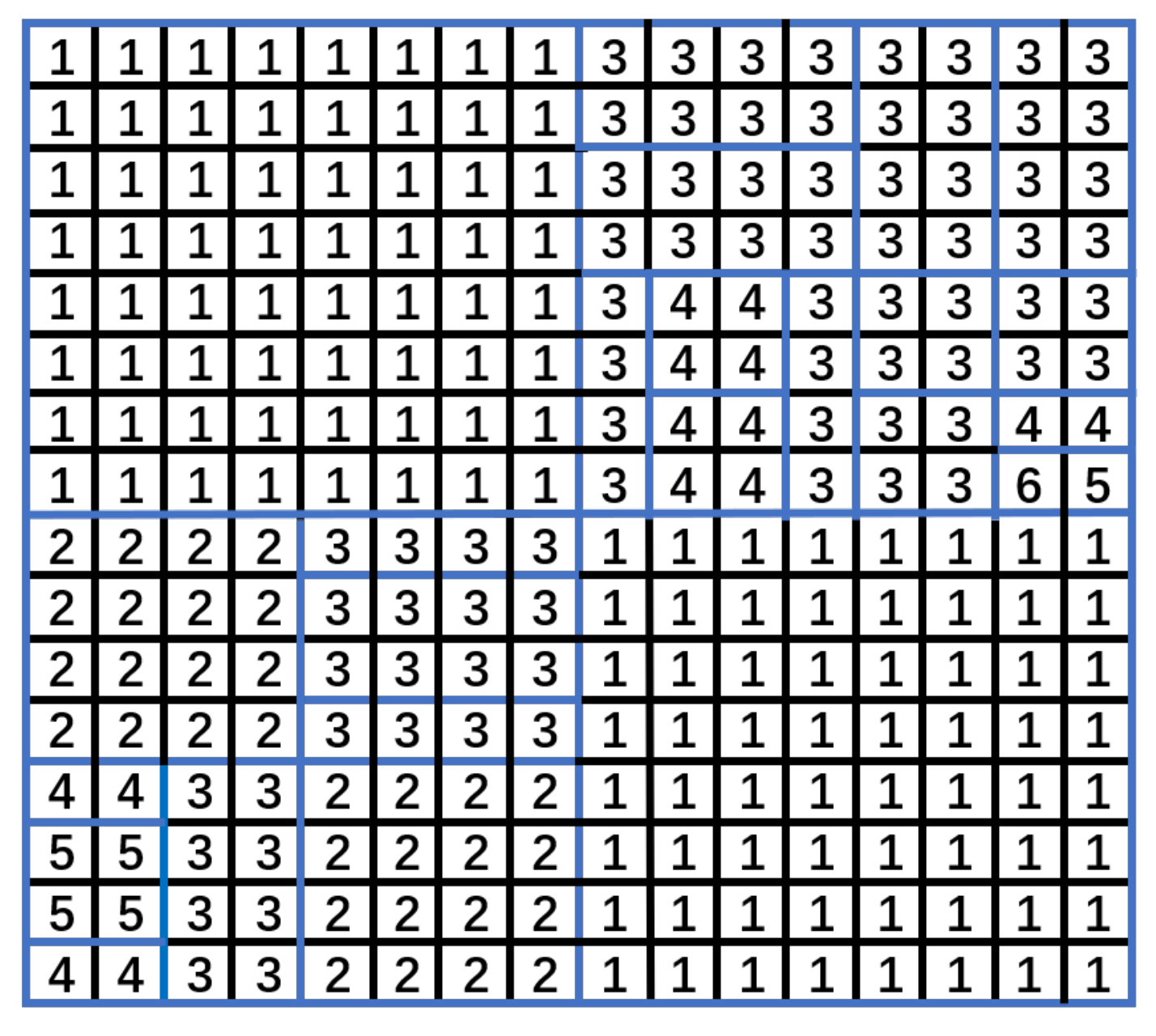}
\end{minipage}
}%
\centering
\caption{ Example of (a) CTU partitioning in VVC and its corresponding (b) depth information and (c) depth map. \label{Fig1}}
\end{figure*}

Nowadays, there are numerous algorithms developed to low complexity intra or inter prediction for HEVC, including non-learning and learning-based algorithms. There are still few contributions on complexity optimization of VVC.

\subsection{Low-Complexity HEVC Prediction without Learning}

Researchers have developed various algorithms aiming at efficient CU partitioning on the HEVC standard \cite{004,0614,005,006,007,0615,008,009,010,011,012,0616,0617,0618,2612693,2815941,2356795}.
Gu {\it et al}. \cite{004} proposed an adaptive approach to boost intra prediction of HEVC, where the intra angular prediction modes and partition depth were predicted by the distribution of coding residual.
In \cite{0614}, Tsang {\it et al}. proposed to reduce the computational complexity of HEVC-based Screen Content Coding (SCC), in which the intrablock copy mode was also optimized. 
In \cite{005}, the rough mode decision candidate list and rate distortion optimization candidate list are narrowed by the spatial-temporal adjacent Prediction Unit (PU).
Fern{\'a}ndez {\it et al}. \cite{006} employed Parseval's relation and the Mean Absolute Deviation (MAD) of motion component to analyze the spatial and temporal homogeneity of video frames, which was further used in CU partition decision.
Zhang {\it et al}.  \cite{2612693} proposed a probability-based early depth intra mode decision to speed up the time-consuming view synthesis optimization.
Based on the Bayesian decision rule and scene change detection, Kuang {\it et al}. \cite{007} proposed an online-learning method to reduce the computational complexity of encoder.
Goswami {\it et al}. \cite{2815941} used two Bayesian classifiers for early skip decision and CU termination respectively. The Markov Chain Monte Carlo method was applied to determine prior and class conditional probability values at run time.
In \cite{2356795}, Xiong {\it et al}. modeled CU partition as a Markov random field and conducted a maximum a posteriori approach to evaluate the splittings of the CUs.
Zhang {\it et al}. \cite{0615} developed a dynamic HEVC coding scheme that achieves a user-defined complexity constraint by adjusting the depth range for each CTU. 
Duanmu {\it et al}. \cite{008} adopted extensive statistical studies and mode mapping techniques to early terminate the partitioning process of CU.
The work in \cite{009} used the depth information of previously coded CUs to adaptively choose the optimal CU depth range.
Shen {\it et al}. \cite{010} skipped the unnecessary CU depth levels and intra angular prediction modes by using the information of upper depth level CUs and neighboring CUs.
\cite{011} predicted the RD cost of intra angular modes to skip the non-promising modes.
Based on the distortion cost obtained from motion estimation, proposed a CU partitioning algorithm to speed up the coding process. In addition, the researchers have developed more than fast partition in HEVC. In \cite{0616} and \cite{0617}, the fast intra and inter partition methods were extended to the scalable video encoders, while in \cite{0618}, an improved RD model achieved its efficiency in Motion Estimation (ME) by introducing complexity optimization.

\begin{figure*}[h]
	\centering
	\includegraphics[width=1\textwidth]{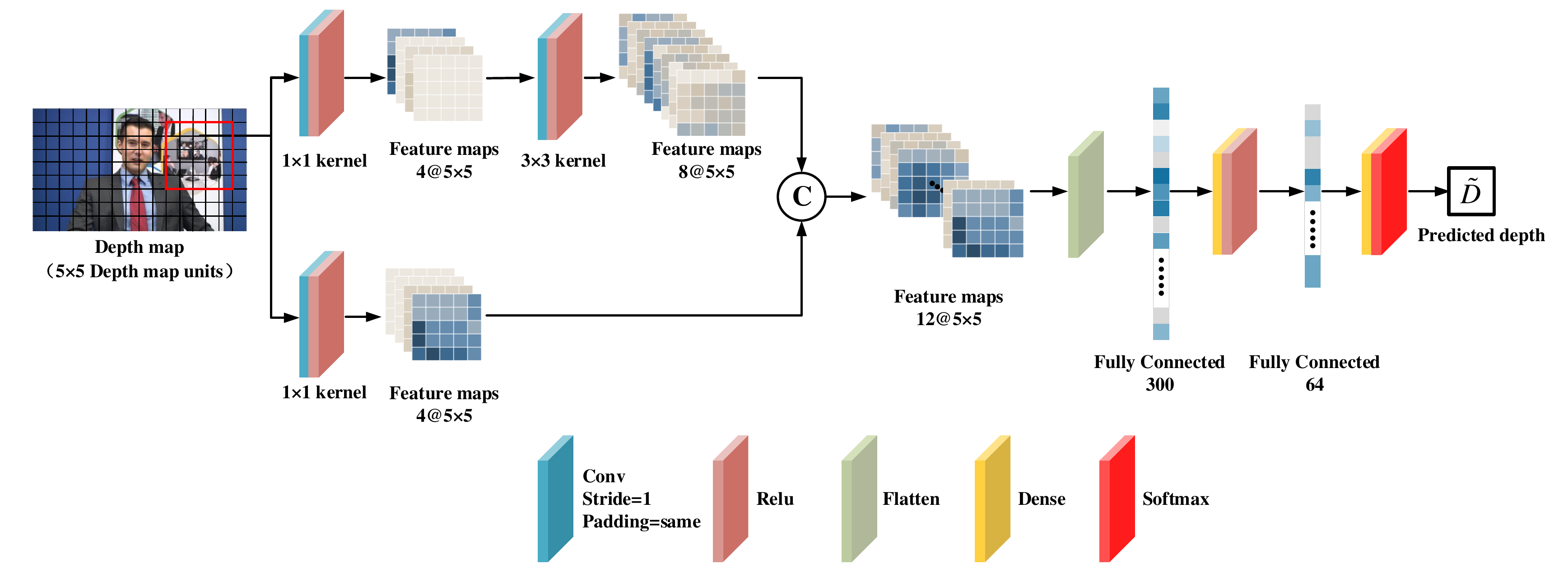}
	\caption{ The architecture of the proposed D-DFF. \label{Fig2}}
	\centering
	\vspace{-1em}
\end{figure*}

\subsection{Low-Complexity HEVC Prediction with Learning}

The application of Machine Learning (ML) has become popular in video coding optimization.
The application of ML-based algorithm brings extra computational load to the encoder, resulting in the increase of coding time.
However, compared with the computational complexity of the encoder, the overhead of ML-based algorithm is negligible \cite{013}.
In \cite{021}, Mercat {\it et al}. demonstrated the superiority of ML-based method to conventional probability-based method in the scenario of fast video coding.
As a popular tool in machine learning, the random forest was demonstrated to effectively predict the intra angular prediction mode by Ryu {\it et al}. \cite{0131}.
In \cite{0619}, Tsang {\it et al}. utilized the random forest to design mode skipping strategy for SCC, in order to reduce the computational complexity of intrablock copy and palette modes. 
Grellert {\it et al}. \cite{014} proposed a Support Vector Machine (SVM) based method to make fast CU partition decision. In this method, the coding flags, coding metrics and motion cues were utilized as the inputs of SVM.
The SVM-based models were also explored in \cite{015,016}, which utilized the average number of contour points in current CU and the image  texture complexity, respectively.
\cite{017} employed the CNN to analyze the luminance components of video frame, which was furthermore employed to determine the depth range of CTU.
In \cite{018}, during the process of fast CU partition decision, CNN was used to decide whether an 8$\times$8 CU needs to be split.
Yeh {\it et al}. \cite{019} designed a simplified CNN structure and a complex CNN structure to enhance the performance of encoder on CPU and GPU, respectively.
Li {\it et al}. \cite{020} established a large-scale database of CU partition and applied a deep CNN to make binary decision for CU partition. This work was further improved in \cite{013} by introducing the Long-Short-Term-Memory (LSTM) network.
In addition, high-level image features have also been adopted in CNN-based fast CU partition.
For example, in \cite{022}, Kuanar {\it et al}. embedded region-wise image features in their CNN model.
Kim {\it et al}. \cite{023} jointly used the image and features extracted from the coding process ({\it e.g.} PU modes, intra angular prediction modes, motion vector) to train the CNN model.

\subsection{Low-Complexity VVC Optimization}
Despite of the above great efforts, they were all designed for the HEVC standard, lacking of application in the newly released VVC. To further exploit the potentiality of video compression, the VVC adopts a more complex QuadTree with nested Multi-type Tree (QTMT) structure, as shown in Fig. 1 (a,b).
As a result, the above approaches cannot be directly applied in VVC and thus new optimization approaches are required.
Cui {\it et al}. \cite{09105861} proposed an early termination algorithm based on horizontal and vertical gradients to pre-determine binary partition or ternary partition.
Variance and Sobel operator were used in \cite{09110597} to terminate further partition of 32$\times$32 CU and decide QTMT structure.
Wang {\it et al}. \cite{08261} employed the residual block as the input of CNN to predict the CU depth and also accelerated the inter coding process.
In \cite{08262}, the inherent texture richness of the image was adopted in CNN to predict the partition depth range of 32$\times$32 block.
\cite{08263} utilized two CNNs to predict boundaries for luma and chroma components, respectively.
Based on random forest, \cite{010101} designed a Quad Tree Binary Tree partitioning scheme to determine the most probable partition modes for each coding block.
\cite{GLCM} extracted texture region features based on gray-level co-occurrence matrix to train random forest classifier, which predicts CU partition. 
To improve the flexible QTMT structure, \cite{09444659} proposed a multi-stage CU partition by a CNN model trained with an adaptive loss function.
Li {\it et al}. \cite{09410290} established two decision models by exploiting the hierarchal correlation of the luma prediction distortion for early skipping Binary Tree (BT) and  Ternary Tree (TT) partition.
To reduce encoding complexity, Wu {\it et al}. \cite{09401614} trained split classifier and splitting directional classifier based on SVM for different sizes of CUs.
In \cite{010102}, a Context-based Ternary Trees Decision (C-TTD) approach was proposed to skip the TT partitions which efficiently reduced the encoding time.
Fu {\it et al}. \cite{030} adopted Bayesian rule to early skip the splitting of CU, in which the split types and intra prediction modes of children CUs were used.
Yang {\it et al}. \cite{0231} proposed a CTU Structure Decision Strategy based on Statistical Learning (CSD-SL) and a fast intra mode decision method for VVC intra coding. Tang {\it et al}. \cite{32} used the block-level based Canny edge detector to extract edge features to skip vertical or horizonal partition modes. According to the edge map, the homogenous CUs can be early terminated.

\section{The Proposed Algorithm}

In the state-of-the-art VVC intra prediction, there are two steps that are successively executed. Firstly, the CTU is iteratively split into a number of CUs with different coding depth. Secondly, in each coding depth, the partition modes with different directions and patterns are thoroughly checked to find the one with the minimum RD cost. Accordingly, we also design a two-stage complexity optimization strategy: the D-DFF to determine the optimal depth and the P-PBE to select the candidate partitions. The chosen depth and partitions are finally used to speed up the CU partition process in VVC intra coding.

\subsection{Intra Depth Prediction with Deep Feature Fusion}

To support videos with higher definition and fidelity, the VVC incorporates larger size and depth of CTU, as compared with HEVC. The complicated CTU structure requires an efficient method to reduce its computational cost. In this work, we divide the CTU into 8$\times$8 blocks and make an attempt to predict the optimal depth of each block. The selection of block size 8$\times$8 is based on a tradeoff between prediction accuracy and coding complexity. As shown in Fig. 1 (c), a CTU of size 128$\times$128 is divided into 16$\times$16 blocks.

To accurately predict the optimal depth values, we need to collect the reference information that is available during video coding. The strong spatial-temporal correlation [15] inspires us to adopt the depth information of spatial-temporal neighboring CUs. For each 8$\times$8 block located at $(x, y, t)$, where $x, y$ and $t$ respectively denote the spatial coordinates and temporal order, we collect depth values of the following blocks:
\begin{equation}\label{equ:Equ1}
\begin{split}
S_{D}=
\begin{cases}
B(x + \Delta x, y + \Delta y , t),  \\  \qquad \quad  \qquad{\text {\text if } \Delta x< 0|| \Delta x =0 \& \Delta y <0}, \ \\
B(x + \Delta x, y + \Delta y , t-1), \qquad  {\text {otherwise}}, \
\end{cases}
\end{split}
\end{equation}
where $\Delta x$ and $\Delta y$ denote the integers ranging from -2 to 2. In other words, we collect the depth of a neighboring block if it has been coded; otherwise, we collect the depth of its co-located block at the previously coded frame.

We extract and fuse the features of these spatial-temporal references with a deep CNN. In particular, the CNN is designed as a lightweight network with multi-scale feature fusion. The lightweight network avoids high computational overhead. The multi-scale feature fusion obtained with different sizes of convolutional kernels provides more accurate classification than single scale features \cite{huangg}. The network infrastructure, training and depth prediction are elaborated as follows.

\emph{\it 1) Network Infrastructure}

The design of proposed model is inspired by the inception module, which was proposed to solve complex image recognition and classification tasks. To avoid large computational overhead, we set the model with reduced parameters and dimensions. As shown in Fig. \ref{Fig2}, the proposed D-DFF consists of three steps: feature extraction, feature concatenation and classification.  

The feature extraction step receives the reference depth map defined in (1). It extracts the features of depth map in two paths: one utilizes a 1$\times$1 convolutional kernel for dimension promotion and a 3$\times$3 convolutional kernel with ReLU for scaled feature extraction; the other utilizes a 1$\times$1 kernel only. With $\Delta x$ and $\Delta y$ ranging from -2 to 2, the size of depth map is 5$\times$5. The two paths output 8 and 4 5$\times$5 feature maps, respectively.  All extracted features are input to the next step for feature fusion.

The feature concatenation step combines all feature maps from the first step and flattens them into a vector. For each depth unit, the 1$\times$1 convolution kernel is equivalent to the unit performing fully connected calculations on all features. Connecting multiple convolutions in series in the same size receptive field benefits the combination of more nonlinear features and extract richer features \cite{huangg}. After this step, the 12 5$\times$5 feature maps are stretched into a vector of length 300.

Finally, the classification step receives the feature vectors and outputs the predicted depth ${\tilde{{D}}(x,y)}$. A neural network with 2 hidden layers and a softmax layer is employed to fulfill this task. Since the intra prediction is performed at CU depth 1 or above, there are only 6 types of output depth from 1 to 6. The depth value with the highest probability is selected as the predicted depth.

\begin{figure}[h]
	
	\centering
	\includegraphics[width=0.45\textwidth]{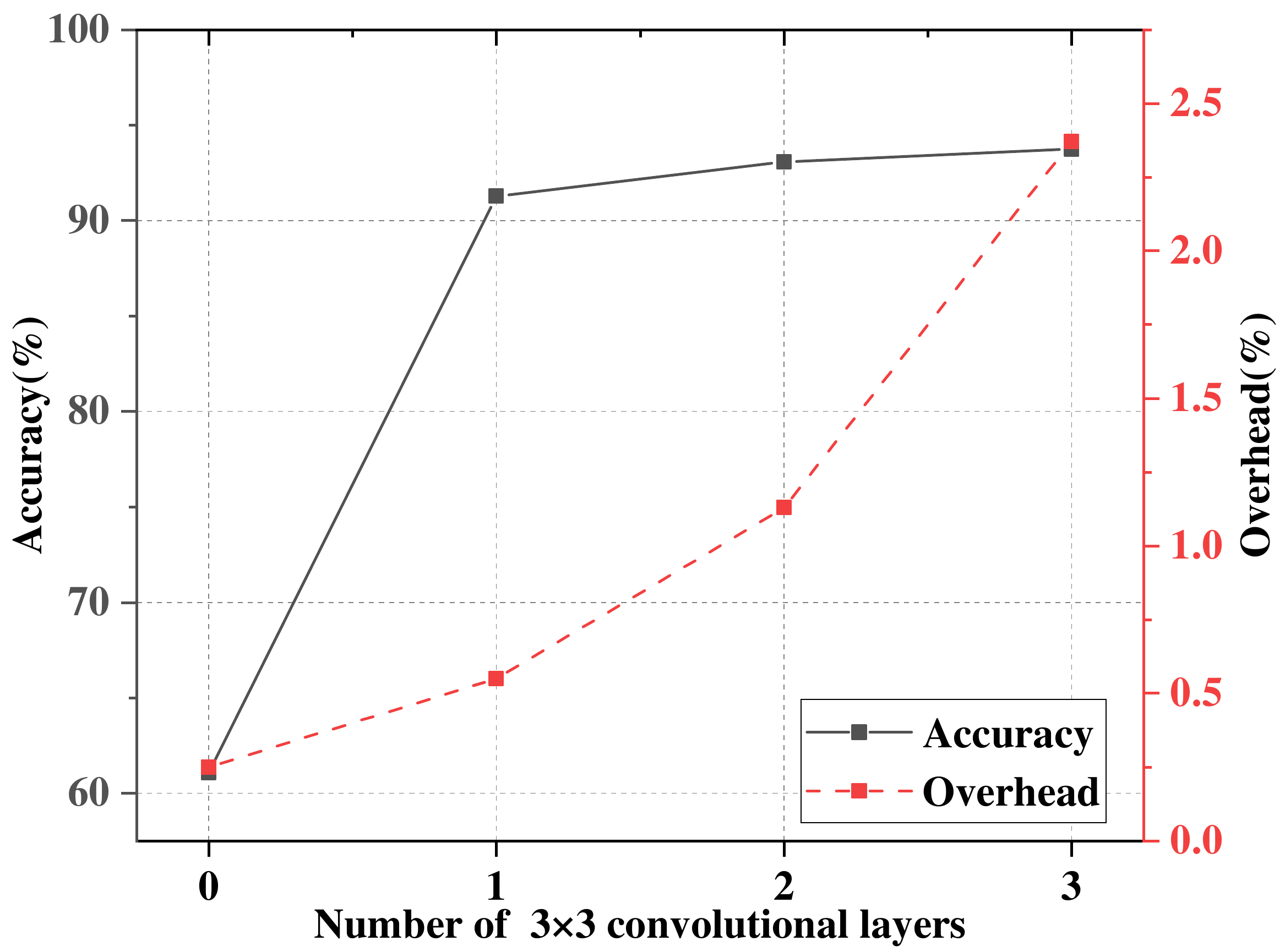}
	\caption{ The model accuracy and computational overhead under different number of 3$\times$3 convolutional layers. \label{accuracy+overhead}}
\end{figure}

Aimed at a tradeoff between model accuracy and complexity, we have optimized all hyper-parameters of this D-DFF model. An important design of this model is its two-path feature extraction to obtain multi-scale features. In this step, the maximum number of convolutional layers influences both the prediction accuracy and the computational overhead of our model. In Fig. \ref{accuracy+overhead}, we present the average accuracy and overhead of our model under different number of 3$\times$3 convolutional layers. Four typical video sequences (\emph{\it BQTerrace}, \emph{\it RaceHorsesC}, \emph{\it BQSquare} and \emph{\it Johnny}) are tested with Qp from 22 to 37. From the figure, the model with one 3$\times$3 convolutional layer achieves a good tradeoff between model accuracy and computational complexity. As a result, we utilize a 1$\times$1 and a 3$\times$3 convolutional layer in the first path. The maximum number of convolutional layers are set as 2. 

\emph{\it 2) Model Training}

\begin{figure}[h]
	\centering
	\includegraphics[width=0.43\textwidth]{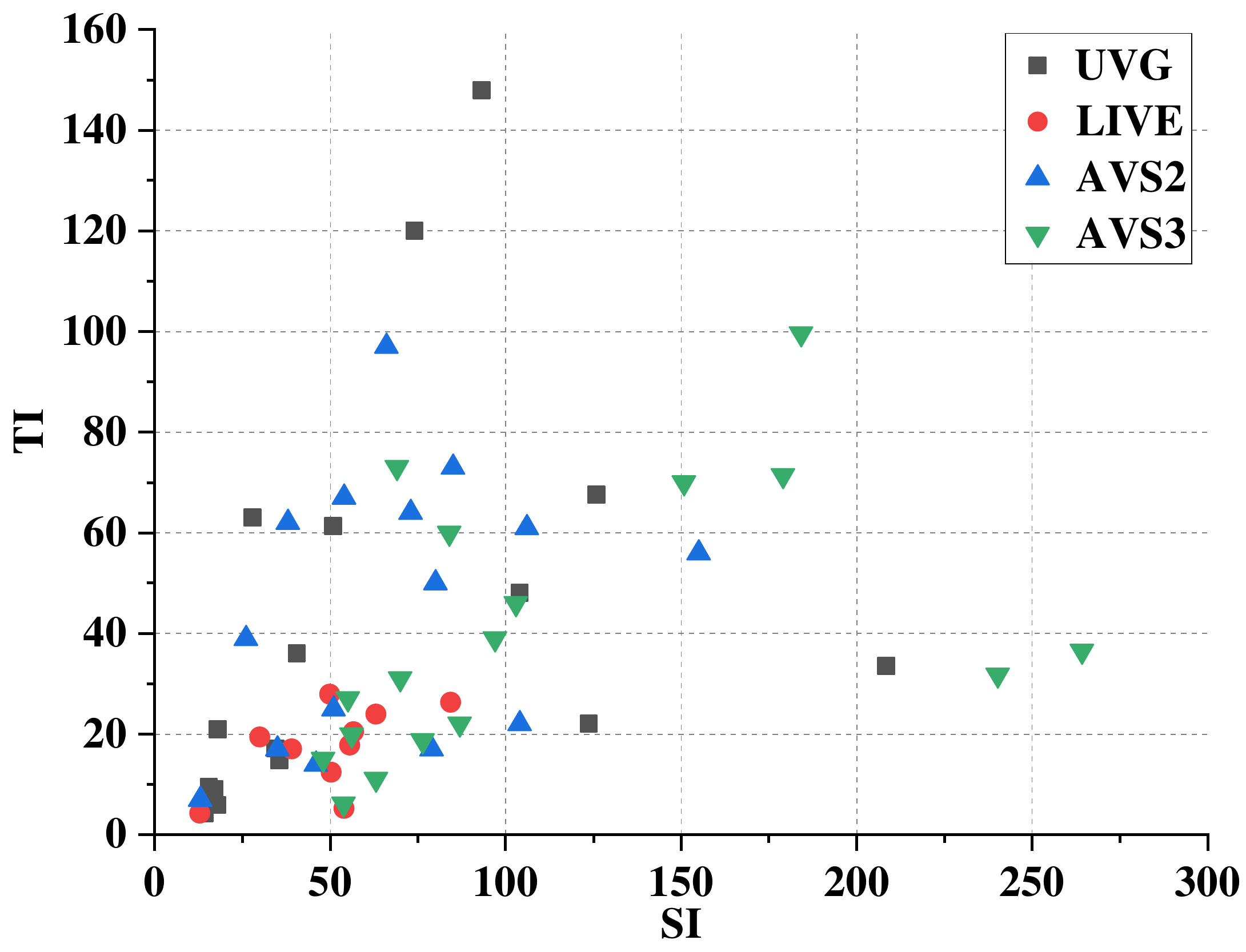}
	\caption{The SI and TI distributions of sequences in Table \ref{tab:curve_plcc} .\label{Figv666}}
	\vspace{-1em}
\end{figure}

We collect 58 video sequences from the LIVE data set \cite{huang17}, the UVG data set \cite{huang14} and the standard sequences of AVS2/AVS3. As shown in Table \ref{tab:curve_plcc}, these sequences cover a large range of resolutions with diversified frame rates and bit depths. As shown in Fig. \ref{Figv666}, they also cover a wide range of Spatial Information (SI) and Temporal Information (TI). These sequences are further compressed by VTM 12.0 with 4 Quantization parameters (Qps) of 22, 27, 32 and 37. Therefore, these compressed sequences are representative to improve the generalization performance of our prediction model. 

\begin{table}[h]
	\vspace{-1em}
	\centering
	\caption{The video sequences for model training and testing}
	\label{tab:curve_plcc}
	\setlength{\tabcolsep}{1mm}{
		\begin{tabular}{cccccc}
			\toprule
			Sources  & Resolution & Number of sequences & Frame rate  & Bit depth 	\\ \midrule
			\multirow{2}{*}{UVG}      
			& 3840$\times$2160 &8 & 50 & 10  \\
			& 1920$\times$1080 &8 & 50 & 8  \\
			\midrule
			\multirow{4}{*}{AVS2}
			& 3840$\times$2160 &2 & 50 & 10  \\
			& 2560$\times$1600 &1 & 30 & 8  \\
			& 1920$\times$1080 &7 & 25 & 8  \\
			& 1280$\times$720 &5 & 60 & 8  \\
			\midrule
			\multirow{4}{*}{AVS3}
			& 3840$\times$2160 &6 & 30 & 10  \\
			& 1280$\times$720 &5 & 30 & 8  \\
			& 720$\times$1280 &3 & 30 & 8  \\
			& 640$\times$360 &3 & 30 & 8  \\
			\midrule
			\multirow{1}{*}{LIVE}
			& 1280$\times$720  & 10 & 24 & 8  \\
			\midrule			
	\end{tabular}}
	\vspace{-1em}
\end{table}

During the compression, we collect all depth values of CUs and reorganize them into pairs of the predicted depths and the corresponding reference depth maps. These data pairs formulate a big dataset which is further divided into a training and a testing set at a 4:1 ratio.  For training, we set the batch size and the iterations as 256 and 128, respectively. We choose Adam optimizer with a learning rate of 0.0001. The cross entropy function, which is popular in classification, is utilized as the loss function.

\begin{figure}[h]
	\vspace{-1em}
	\centering
	\includegraphics[width=0.43\textwidth]{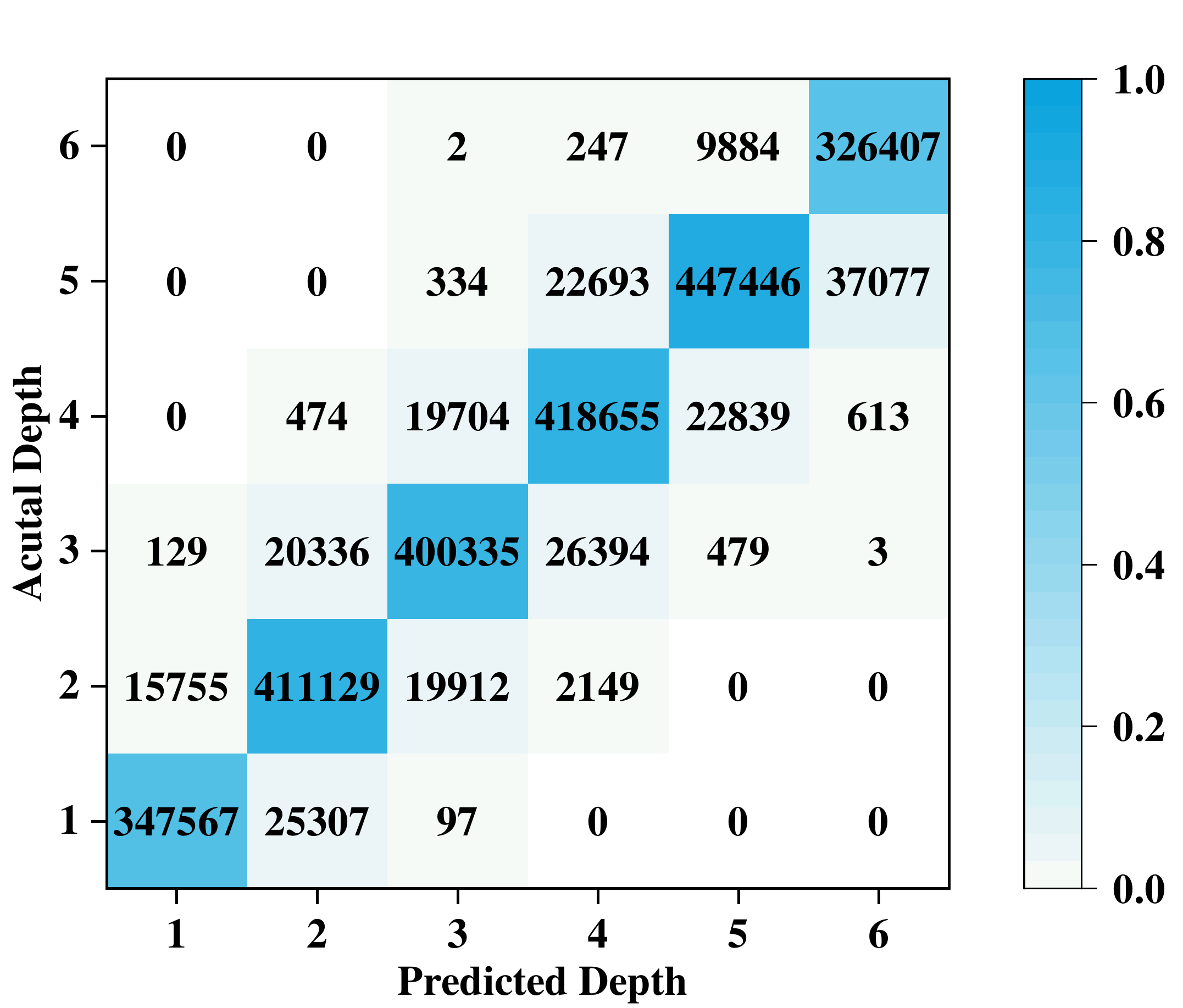}
	\caption{The heat map of our depth prediction model.\label{TMM_Heatmap}}
\end{figure}

Our deep model demonstrates high prediction performance in the testing set. As shown in Fig. \ref{TMM_Heatmap}, $2351539/2575967=91.29\%$ depth values are correctly predicted with another $219901/2575967=8.54\%$ within the error of ${\pm 1}$. As shown in Table \ref{classification_performance}, the proposed model achieves the average precision, recall, specificity and accuracy of 0.914, 0.917, 0.983 and 0.971, respectively. These results demonstrate the effectiveness of our model, especially when considering its lightweight infrastructure. 

\begin{table}[h]
	\vspace{-1em}
	\centering
	\caption{The classification performance of our depth prediction model}
	\label{classification_performance}
	\setlength{\tabcolsep}{3mm}{
		\begin{tabular}{ccccc}
			\toprule
			Depth & Precision & Recall & Specificity & Accuracy	\\ \midrule
			1      & 0.956 & 0.932 & 0.993 & 0.984 \\
			2      & 0.900 & 0.916 & 0.978 & 0.967 \\
			3      & 0.909 & 0.894 & 0.981 & 0.966 \\
			4      & 0.890 & 0.906 & 0.976 & 0.963 \\
			5      & 0.931 & 0.882 & 0.984 & 0.964 \\ 
			6      & 0.896 & 0.970 & 0.983 & 0.981 \\ \midrule
			Average                     & 0.914 & 0.917 & 0.983 &  0.971 \\ \midrule
			
	\end{tabular}}
	\vspace{-1em}
\end{table}

\emph{\it 3) The Depth Prediction}

Our deep model achieves a tradeoff between the prediction accuracy and the computational overhead. However, in video coding optimization, a tradeoff shall be achieved between the finally achieved RD performance and the overall computational complexity. Hence, there exists a gap between the two tradeoffs. The first tradeoff chooses the most probable depth, which may still lead to a small portion of incorrect prediction. These incorrect predictions may accumulate to a considerable quantity during video coding and further result in an intolerable increment of RD cost. To avoid this error propagation, we employ a conservative strategy as follows:
\begin{equation}
\label{huang1}
\begin{aligned}
\begin{split}
{\hat{D}(x,y)}=&{\tilde{D}(x,y)}+ \\
&\max\{\lfloor \frac{1}{{K}_{D}} \sum\limits_{k=1}^{{K}_{D}} ({D}_{k}(x,y)-\tilde{{D}_{k}}(x,y)) + \frac {1}{2} \rfloor,0\} ,
\end{split}
\end{aligned}
\end{equation}
where $\hat{D}(x,y)$ represents the adjusted depth. $K_D=25$ denotes the number of reference blocks to predict the current depth. $D_k(x,y)$ and $\tilde D_k(x,y)$ represent the final and predicted depth of $k$-th reference block. Through (2), we are able to estimate a depth offset from its neighboring blocks. This method effectively reduces the RD loss caused by accumulated prediction error and further improves the robustness of our D-DFF model.

We can derive the optimal depth of each CU based on the adjusted depth $\hat D(x,y)$. In a CU with multiple 8$\times$8 blocks, its optimal coding depth is estimated as 
\begin{equation}\label{equ:Equ51}
{D}_{o}= \max \limits_{\{x,y\}\in \text{CU}}\{\hat{D}(x,y)\}.
\end{equation}

During coding, the current CU is iteratively split until its optimal depth. The coding depth larger than the optimal depth is skipped to save the coding time.

\subsection{Intra Partition Mode Prediction with Probability Estimation}

In CTU coding, the splitting process is iteratively performed until the optimal depth $D_o$ of each CU. For each depth less than $D_o$, the CU traverses 5 possible partition modes, including Quadtree (QT) partition, Vertical Binary Tree (BTV) partition, Horizontal Binary Tree (BTH) partition, Vertical Ternary Tree (TTV) splitting and Horizontal Ternary Tree (TTH) splitting. To further skip the unnecessary coding modes, we need to predict the probability of all mode partitions. As indicated above, the spatial-temporal correlation and reference mode partitions could be investigated for partition mode prediction.

Denote a CU located at $(x, y, t)$ as $U(x, y, t)$, whose size is larger than 4$\times$4 in VVC. We set its reference set as
\begin{equation}\label{equ:Equ2}
\begin{split}
S_{P}=
\begin{cases}
U(x + \Delta x, y + \Delta y , t||t-1),  \\ 
\qquad \quad  \qquad{\text {\text if } \Delta x< 0|| \Delta x =0 \& \Delta y <0}, \ \\
U(x + \Delta x, y + \Delta y , t-1), 
\qquad  {\text {otherwise}}, \
\end{cases}
\end{split}
\end{equation}
where $\Delta x$ and $\Delta y$ ranges from -1 to 1. The reference set (4) is similar to (1) but with two differences. Firstly, we collect the partitions of top and left CUs in both the current and the left frames. Previous works showed that all these CUs have high partition correlations to the current CU \cite{0251}, which inspires us to add all these CUs in the reference set $S_P$. Meanwhile, the probability-based partition prediction does not require a neat matrix for convolution, which also allows us to add these CUs. Secondly, the ranges of $\Delta x$ and $\Delta y$ are reduced. Without the convolutional operation, a smaller but effective reference set is more practical. 

Let $R$ denote the set that consists of all best partitions modes in the reference CU set $S_P$, $R=\{{\rm bm}_U|U\in S_p\}$. For a partition mode $M$, its probability to be selected as the best partition mode can be estimated as:
\begin{equation}\label{equ:Equ9}
\begin{aligned}
P(\text{bm}=M) =
P(\text{bm}=M|M\in R)P(M\in R)  \\
+P(\text{bm}=M|M\notin R)P(M\notin R),
\end{aligned}
\end{equation}
where $P(M \in R)$ and $P(M \notin R)$ are posterior probabilities that are either 0 or 1. Experimental results in Table III also shows that $P({\rm bm}=M | M \in R) \gg P({\rm bm}=M | M \notin R) $, where four typical sequences from the standard sequences of AVS2/AVS3 with different resolutions and Qps from 22 to 37 are examined. Here it is reasonable to set
\begin{equation}\label{equ:Equ11}
P(\text{bm}=M|M\notin R) \approx 0,
\end{equation}
which simplifies our derivation. The other part of (5) can be easily derived as
\begin{equation}\label{equ:Equ13}
P(\text{bm}=M|M\in R)= \frac {\int_{U\in S_P} {P(bm_{U}=M)}}{\int_{m\in R} \int_{U\in S_P}{P(bm_{U}=m)}},
\end{equation}
where $S_p$ and $R$ denote the reference CU and partition sets, as indicated above. In practice, the probabilities in (6) are estimated with the number of occurrences in coding history:
\begin{equation}\label{equ:Equ10}
P(\text{bm}=M|M\in R)= \frac {\sum\limits_{U\in S_P} {N(bm_{U}=M)}}{\sum\limits_{m\in R} \sum\limits_{U\in S_P}{N(bm_{U}=m)}},
\end{equation}
where $N(\cdot)$ indicates the statistical frequency of an event. Experimental results in the following demonstrate the effectiveness of our method.

\begin{table}[h]
	\small
	\centering
	\caption{The average probabilities to be the best partition mode}
	\label{TableV7_2}
	
	\begin{tabular}{ccc}
		\toprule
		\multirow{2}{*} {Sequences}   & \multicolumn{2}{c}{Average Probabilities(\%)}\\
		\cmidrule(r){2-3}
		& $P(\text{bm}=M|M\in R)$     & $P(\text{bm}=M|M\notin R)$\\
		
		\midrule
		\emph{Pedestrian\_area}                 &36.74                 &8.16    \\
		\emph{Yellowflower}                     &36.36                 &7.14    \\
		\emph{Night}                            &36.24                 &4.65    \\
		\emph{ShuttleStart}                     &36.88                 &6.94    \\
		
		\bottomrule
	\end{tabular}
	\vspace{-1em}
\end{table}

After obtaining the probability of each partition mode, we sort the partition modes which belong to $R$ according to their descending probabilities and add the other modes (not in $R$) after them. When the RD cost of current partition mode ($J_{\rm cur}$) is larger than the minimum RD cost obtained
so far ($J_{\rm min}$) , we skip the untested partitions to save the total encoding time. The flowchart of the proposed intra partition mode prediction, namely P-PBE, is shown in Fig. 6.

\begin{figure}[h]
	\vspace{-0.5em}
	\centering
	\includegraphics[scale=0.6]{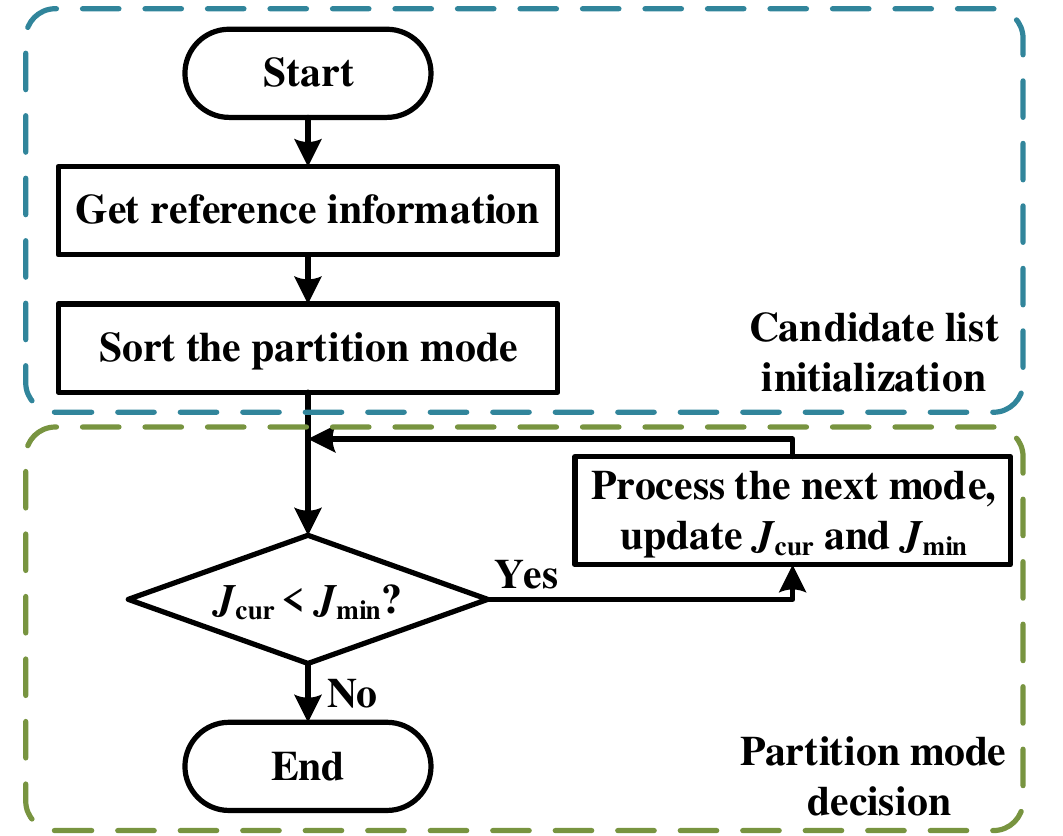}
	\caption{The proposed P-PBE method.\label{Fig6}}
	\vspace{-1em}
\end{figure}

\begin{table*}[htb]
	\small
	\renewcommand\arraystretch{0.6}
	\renewcommand\tabcolsep{3.8pt}
	\centering
	\caption{Results of the proposed algorithm compared with the state-of-the-arts}
	\label{Table4}
	\begin{tabular}{cccccccccccc}
		\toprule
		\multirow{2}{*}{\bfseries{Class}} & \multirow{2}{*}{\bfseries{Sequences}} & \multicolumn{2}{c}{\bfseries{C-TTD \cite{010102}}} & \multicolumn{2}{c}{\bfseries{Fu's \cite{030}}} & \multicolumn{2}{c}{\bfseries{CSD-SL \cite{0231}}} &\multicolumn{2}{c}{\bfseries{Tang's \cite{32}}} & \multicolumn{2}{c}{\bfseries{Proposed}}  \\
		\cmidrule(r){3-4} \cmidrule(r){5-6} \cmidrule(r){7-8} \cmidrule(r){9-10} \cmidrule(r){11-12}
		~& ~&\bfseries{ATS($\%$)}  &\bfseries{BDBR($\%$)}    &\bfseries{ATS($\%$)}    &\bfseries{BDBR($\%$)}      & \bfseries{ATS($\%$)} &\bfseries{BDBR($\%$)} &\bfseries{ATS($\%$)} &\bfseries{BDBR($\%$)} 
		&\bfseries{ATS($\%$)} &\bfseries{BDBR($\%$)} \\
		
		\midrule
		\multirow{4}{*}{\shortstack{A1}}
		& \emph{Tango}             		 	&35.32&0.80	&49.22 &1.53	&46.81&0.77                &41.78 &0.69                &67.02 &1.33            \\
		~ & \emph{FoodMarket4}       		 	&29.39&0.77	&50.29 &0.95	&50.88&0.53                &32.08 &0.53               &53.17&0.97                     \\
		~ & \emph{Campfire}          		  	&35.32&0.61	&53.51 &0.90	&42.11&0.83                &34.37 &0.98              &57.32&1.56                      \\
		\cmidrule(r){2-12}
		~ &{\bfseries{Average}}
		&\bfseries{33.34}&\bfseries{0.73}   &\bfseries{51.01}&\bfseries{1.13}
		&\bfseries{46.60}&\bfseries{0.71}   &\bfseries{36.08}&\bfseries{0.73}   &\bfseries{59.17}&\bfseries{1.29}       \\
		\midrule
		\multirow{4}{*}{\shortstack{A2}}
		& \emph{CatRobot}             		  	&28.66&1.04	&36.84&1.26	&43.64&0.84                &38.48 &0.92             &63.18&1.63                              \\
		~ & \emph{DaylightRoad2}        		  	&30.55&1.20	&40.13&1.29	&58.75&0.41                &39.04 &0.68              &62.88&1.23                   \\
		~ & \emph{ParkRunning3}         		  	&33.50&0.54	&48.03&0.44	&51.76&0.62                &34.66 &0.58             &59.52&0.88                        \\
		\cmidrule(r){2-12}
		~ &{\bfseries{Average}}
		&\bfseries{30.90}&\bfseries{0.93}   &\bfseries{41.67}&\bfseries{0.99}
		&\bfseries{51.38}&\bfseries{0.63}   &\bfseries{37.39}&\bfseries{0.73}   &\bfseries{61.86}&\bfseries{1.25}       \\
		\midrule
		\multirow{6}{*}{\shortstack{B}}
		& \emph{BasketballDrive}  &25.68&0.90	&45.37&0.85	&58.43&1.58                &45.61 &0.93             &60.35&1.53                             \\
		~ & \emph{BQTerrace}        &27.42&0.73	&47.28&0.85	&56.22&2.14                &36.64 &0.84             &56.19&1.16                  \\
		~ & \emph{Cactus}           &27.55&0.75 	&48.91&1.29	&59.94&2.06                &39.51 &0.91             &62.98&1.78                       \\
		~ & \emph{Kimono}           &28.17  &0.82  &47.39 &1.10       &57.95 &1.42              &32.69 
		 &0.44 
		             &67.04 
		             &0.93 
		                                   \\
		~ & \emph{ParkScene}        &25.60  &0.76  &46.98 &0.90           &54.88 
		&1.18 
		                &41.95 
		                 &0.56 
 &59.66 
 &1.47                        \\
		\cmidrule(r){2-12}
		~ &{\bfseries{Average}}
		&\bfseries{26.89}&\bfseries{0.79}
		&\bfseries{47.19}&\bfseries{1.00}
		&\bfseries{57.48}&\bfseries{1.68}&\bfseries{39.28}&\bfseries{0.73}   &\bfseries{61.25}&\bfseries{1.37}       \\
		\midrule
		\multirow{4}{*}{\shortstack{C}}
		& \emph{BasketballDrill}  		  	&29.12 
		&1.19 
		&41.39 
		&1.81 
		&46.08 
		&1.89 
		      &29.28 
		       &1.21 
		         &48.91 
		         &1.99 
 \\
		~ & \emph{BQMall}           		  	&30.98 
		&1.05 
			&40.64 
			&0.92 
				&50.28 
				&1.86 
				    &37.73 
				     &0.91 
				      &51.22 
				      &2.02 
           \\
		~ & \emph{PartyScene}       		  	&39.17 
		&0.61 
			&40.14 
			&0.40 
				&49.06 
				&0.76 
				 &35.68 
				  &0.41 
				   &49.86 
				   &0.87 
				                        \\
		~ & \emph{RaceHorsesC}      		  	&38.52 
		&0.64 
			&46.38 
			&0.75 
			&44.96 
			&1.06 
			&32.62 
			&0.61 
			&49.98 
			&1.27 
			                        \\
		\cmidrule(r){2-12}
		~ &{\bfseries{Average}}
		&\bfseries{34.45}&\bfseries{0.87}&\bfseries{42.14}&\bfseries{0.97}
		&\bfseries{47.59}&\bfseries{1.39}&\bfseries{33.83}&\bfseries{0.79}   &\bfseries{49.99}&\bfseries{1.54}       \\
		\midrule
		\multirow{5}{*}{\shortstack{D }}
		& \emph{BasketballPass}   		  	&28.49 
		&0.84 
		&40.10 
		&0.83 
		&48.93 
		&2.53 
		&30.78 
		&0.51 
		&43.62 
		&1.54 
		                     \\
		~ & \emph{BlowingBubbles}   		  	&28.31 
		&0.71 
		&43.14 
		&0.84 
		&35.31 
		&0.72 
		&30.88 
		&0.26 
		&39.74 
		&0.91 
		                     \\
		~ & \emph{BQSquare}         		  	&35.89 
		&0.51 
		&44.20 
		&0.56 
		&44.16 
		&0.79 
		&27.53 
		&0.23 
		&45.31 
		&0.79 
		\\
		~ & \emph{RaceHorses}       		  	&30.49 
		&0.77 
		&45.36 
		&0.97 
		&47.08 
		&0.85 
		&26.63 
		&0.33 
		&48.93 
		&1.09 
		 \\
		\cmidrule(r){2-12}
		~ &{\bfseries{Average}}
		&\bfseries{30.79}&\bfseries{0.71}&\bfseries{43.20}&\bfseries{0.80}
		&\bfseries{43.87}&\bfseries{1.22}&\bfseries{28.95}&\bfseries{0.33}   &\bfseries{44.40}&\bfseries{1.08}       \\
		\midrule
		\multirow{4}{*}{\shortstack{E}}
		& \emph{FourPeople}       		  	&29.00 
		&1.08 
		&41.79 
		&1.60 
		&58.85 
		&2.82 
		&44.29 
		&1.31 
		&58.45 
		&1.97
		 \\
		~ & \emph{Johnny}           		  	&31.59 
		&0.99 
		&40.11 
		&1.38 
		&55.63 
		&3.32 
		&39.96 
		&1.21 
		&59.37 
		&2.05 
		\\
		~ & \emph{KristenAndSara}   		  	&32.18 
		&0.91 
		&37.24 
		&1.14 
		&60.20 
		&2.70 
		&40.14 
		&1.02 
		&58.21 
		&1.90 
		 \\
		\cmidrule(r){2-12}
		~ &{\bfseries{Average}}
		&\bfseries{30.92}&\bfseries{1.00}&\bfseries{39.71}&\bfseries{1.37}
		&\bfseries{58.23}&\bfseries{2.95}&\bfseries{41.46}&\bfseries{1.18}   &\bfseries{58.67}&\bfseries{1.97}       \\
		\midrule
		\multicolumn{2}{c}{\bfseries{Total Average}}
		&\bfseries{30.95}&\bfseries{0.83}&\bfseries{44.29}&\bfseries{1.03}
		&\bfseries{50.99}&\bfseries{1.44}&\bfseries{36.01}&\bfseries{0.73}   &\bfseries{55.59}&\bfseries{1.40}       \\
		\bottomrule
	\end{tabular}
\end{table*}

\subsection{The Overall Algorithm}

The overall algorithm is proposed as a two-stage method with D-DFF and P-PBE, which address the complexity optimizations at depth and partition levels, respectively. In particular, our method exploits the coding information of reference blocks and CUs, which are not always available for all cases. For D-DFF, an incomplete reference set $S_D$ cannot be processed by convolutions. In such case, the optimal depth $D_o$ of a CU is set as the same as that in its previously co-located CU. For P-PBE, an incomplete reference set still works to derive the probabilities. Therefore, the derivations in (6,7,8) can be processed during coding. 

We summarize the steps of our algorithm as follows:

\newcounter{step}
\begin{list}{\bfseries \upshape \small Step \arabic{step}.}
{
\usecounter{step}
\setlength{\leftmargin}{1.15cm} \setlength{\labelwidth}{1.0cm}
\setlength{\labelsep}{0.10cm} }
\item For the first frame of each video sequence, encode the CUs with original encoder. Go to Step 2.
\item Obtain the reference depth map $S_D$ and predict the depth with our CNN model. For each CU in this CTU determine its optimal depth $D_o$ with (3). Go to Step 3.
\item Check the CTU coding iteratively. Once the current depth of CU exceeds its optimal depth, terminate the encoding process of this CU; otherwise, go to Step 4.
\item For each coding depth, initialize the reference partition set $S_P$. Sort the partition modes in $R$ by (8) and add the remaining modes after them. Go to Step 5.
\item Check the partition modes sequentially. If the current RD cost $J_{\rm cur}$ is larger than the minimum RD cost obtained so far $J_{\rm min}$, terminate the partition selection process and go to Step 4 for the next CU. If all CUs within this CTU has been checked, go to Step 2 to process the next CTU.
\end{list}

\section{Experimental Results}



\begin{figure}[h]
	\vspace{-1em}
	\centering
	\includegraphics[width=0.48\textwidth]{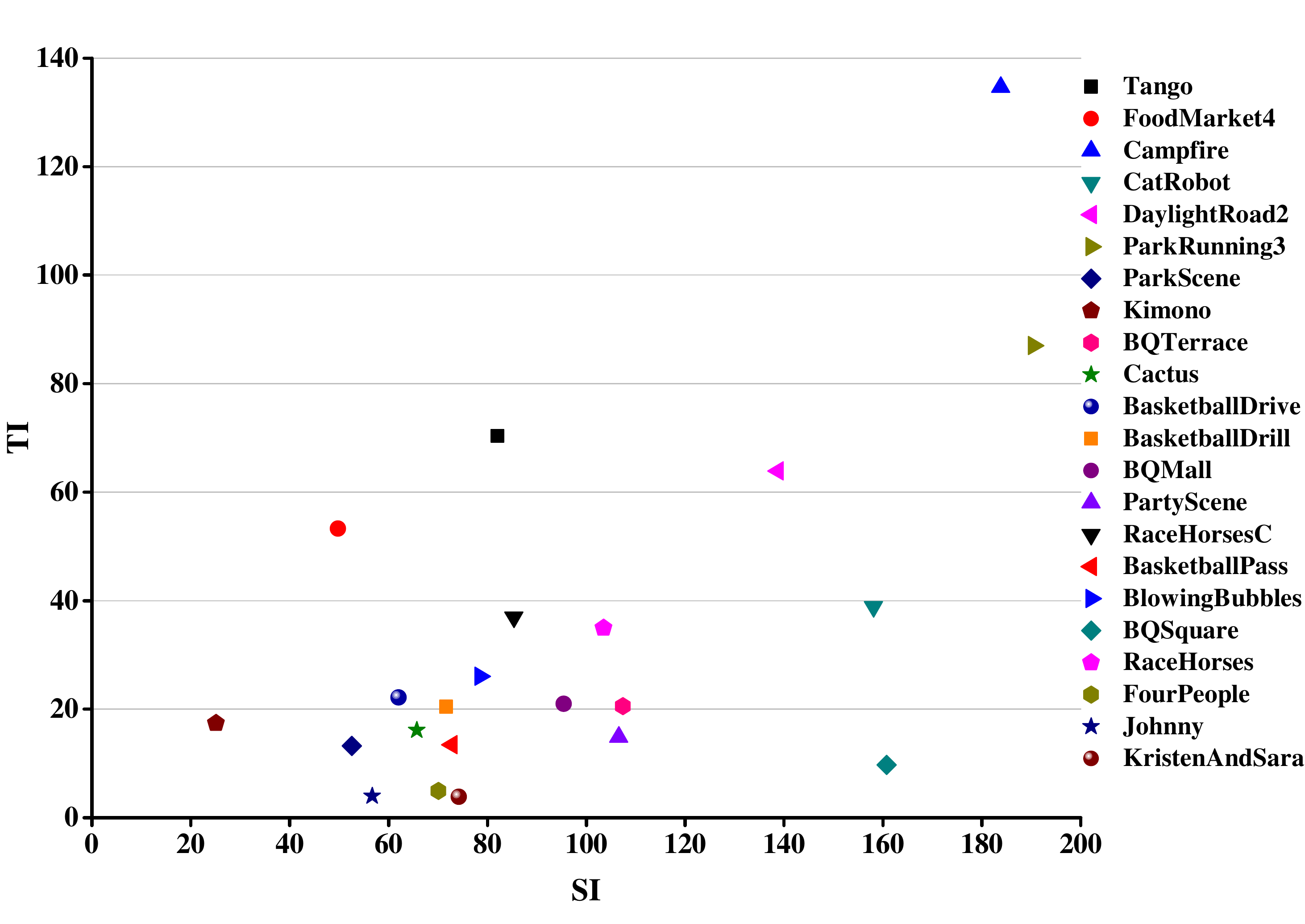}
	\caption{The distributions of SI and TI for all CTC sequences.\label{Figv66}}
\end{figure}

To verify the effectiveness of our proposed algorithm, we implement it on the VTM-12.0 platform \cite{vtm12} under JVET Common Test Condition (CTC) \cite{025} with \emph{\it ALL-INTRA} configurations.
The CTC provides 6 groups of video sequences, including A1 (3840$\times$2160), A2 (3840$\times$2160), B (1920$\times$1080), C (832$\times$480), D (416$\times$240) and E (1280$\times$720). These sequences are with a variety of Spatial Information (SI) and Temporal Information (TI) values \cite{037}, as demonstrated in Fig. \ref{Figv66}. Therefore, the comprehensive examinations are sufficiently representative to examine the proposed algorithm. 

We compare the proposed algorithm and state-of-the-arts including C-TTD \cite{010102}, Fu's \cite{030} , CSD-SL \cite{0231} and Tang's \cite{32}. The evaluation criteria include the Bj{\o}ntegaard Delta Bit Rate (BDBR) ($\%$) and the Average Time Saving (ATS) ($\%$) under four Qps (22, 27, 32 and 37), in which BDBR is defined in \cite{038}, and ATS is defined as
\begin{equation}\label{equ:Equ11}
ATS= \frac {1}{4} \times \sum\limits_{i\in F} \frac{T_{\text{ori}}(i)-T_{\text{pro}}(i)}{T_{\text{ori}}(i)}\times 100 \%,
\end{equation}
where ${T_{\text{ori}}(i)}$ and ${T_{\text{pro}}(i)}$ respectively denote the encoding time of the original encoder and a proposed algorithm when encoding Qp is $i$. $F$ represents the Qp set above. An algorithm with lower BDBR or higher ATS is considered as a superior algorithm.

\subsection{Comparison results with other benchmarks}

The comparison results are shown in Table \ref{Table4}. The four comparison algorithms are implemented in different VTM versions. For fair comparison, we transplant their algorithms to VTM-12.0. All sequences from standard set are examined. 

From Table \ref{Table4}, our scheme has superior computational complexity reduction compared to the other algorithms.
The maximum time reduction occurs at \emph{\it Kimono} sequence. \emph{\it Kimono} has large smooth area with slow motion, thus few CTUs in \emph{\it Kimono} need to be deeply split in our algorithm. On the contrary, the least time reduction is achieved when encoding \emph{\it BlowingBubbles}. The reason mainly lies in its complex textures.
In addition, videos with intense motion ({\it e.g.} \emph{\it BasketballDrill}) have low correlation between neighboring CU depth map units. It reduces the prediction accuracy of the proposed algorithm, resulting in large coding bit-rate increase.

The other three algorithms, C-TTD \cite{010102}, Fu's \cite{030} and Tang's \cite{32}, achieve lower BDBR losses. However, their average encoding time savings are limited at 30.95$\%$, 44.29$\%$ and 36.01$\%$, respectively. On the other hand, the proposed method reduces the computational complexity of encoder by 55.59$\%$ with 1.40$\%$ BDBR increment. Thus, the proposed method further accelerates the encoding speed with acceptable coding loss.

Compared with CSD-SL \cite{0231}, our algorithm improves 0.04\% in BDBR and 3.60\% in ATS. Firstly, for Class A1 and A2, our method focuses on improving ATS, which can save up to 20.21\% more coding time than CSD-SL. Secondly, for the Class B, our method outperforms in both ATS and BDBR. Thirdly, for the Class E, our method can save 0.98\% more bit rate under the same coding time. In addition, our algorithm also has the following advantages. \emph{\it i)} It extracts relatively concise features and uses relatively less network parameters, without tedious calculations. \emph{\it ii)} It utilizes a uniform network model for different partition modes, which is easier to be implemented than other methods such as CSD-SL. \emph{\it iii)} It has more time saving in HD and UHD videos ({\it e.g.} Class A1, Class A2 and Class B), which is favorable under the booming of HD and UHD in videos.


The performance of our proposed algorithm are benefited from two modules. Firstly, the D-DFF effectively terminates the iterative partition process, by utilizing a lightweight CNN to predict an optimal coding depth. Secondly, at each depth, the P-PBE effectively skips the unnecessary partition methods, by exploiting the neighboring coding information. In conclusion, the proposed algorithm achieves better encoding time reduction compared with other four algorithms with slight BDBR loss and simple implementation structure.
\begin{figure}[t]
	\centering
	\subfigure[\label{Fig6a}]{
		\begin{minipage}[ht]{0.6\linewidth}
			\centering
			\hspace*{-1.75cm}
			\includegraphics[width=1.57\textwidth]{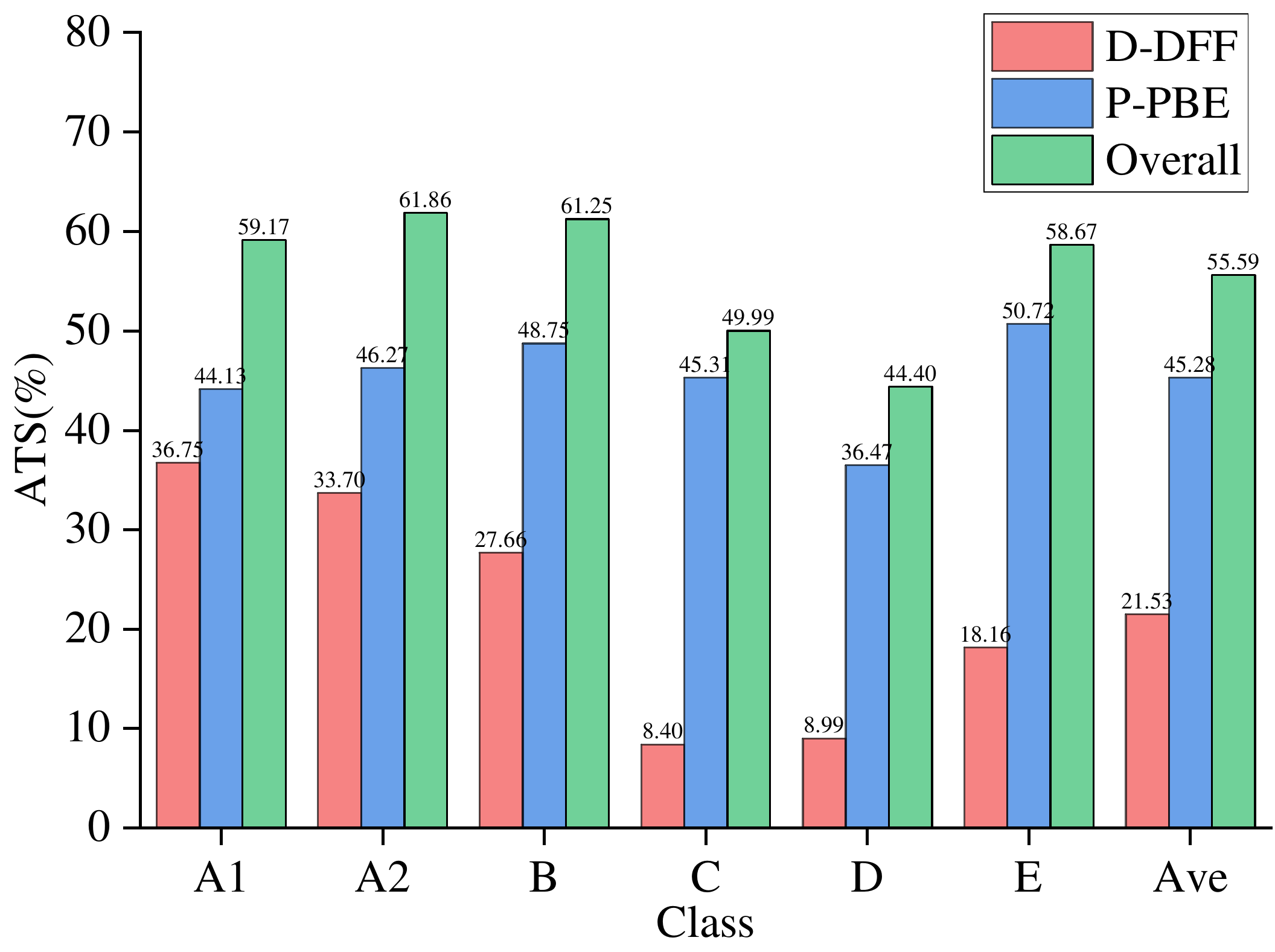}
		\end{minipage}%
	}%
	
	\subfigure[\label{Fig6b}]{
		\begin{minipage}[t]{0.6\linewidth}
			\centering
			\hspace*{-1.7cm}
			\includegraphics[width=1.55\textwidth]{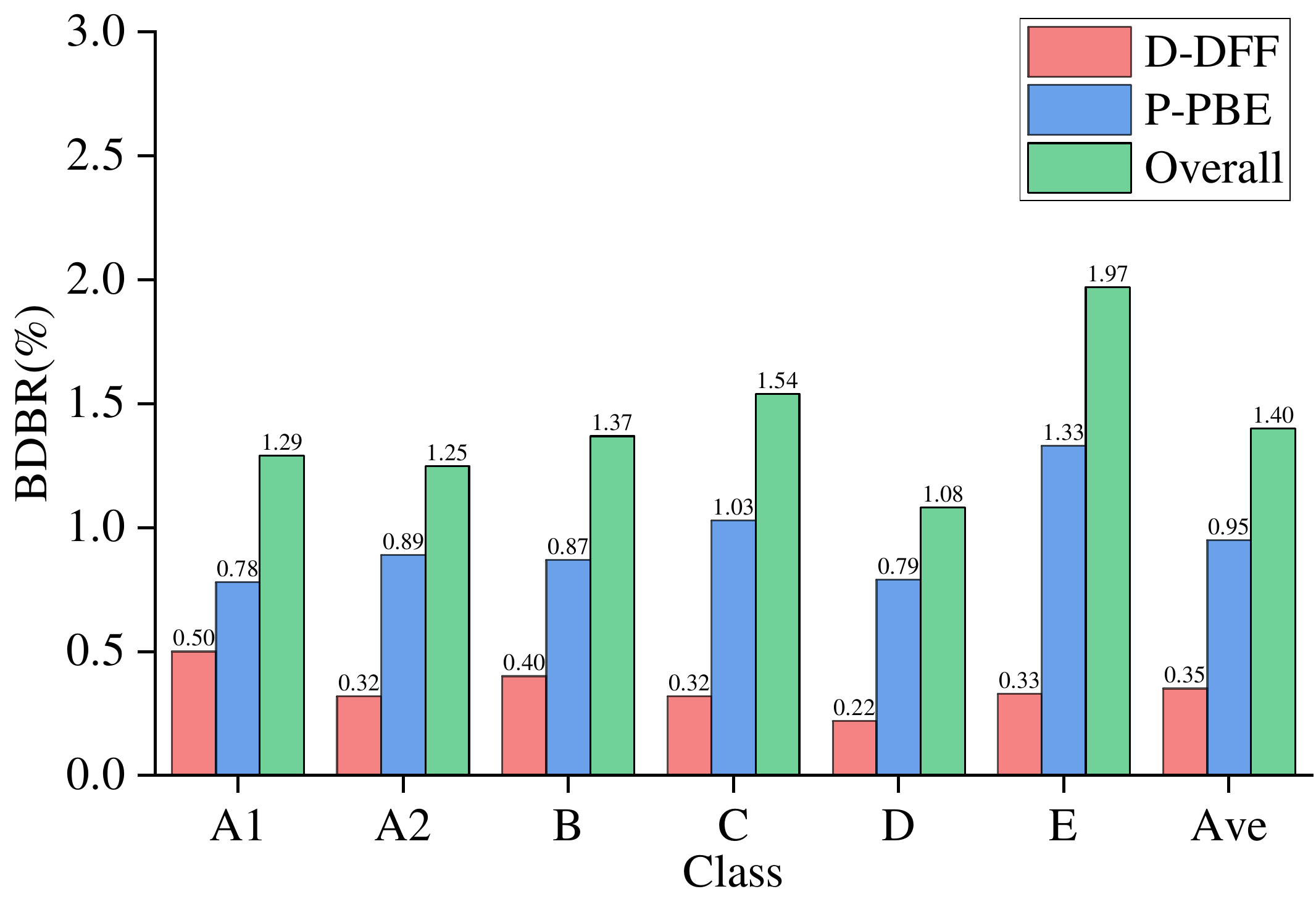}
		\end{minipage}%
	}%
	\centering
	\caption{ Results of ablation experiments in terms of (a) ATS and (b) BDBR.  \label{Fig66}}
	\vspace{-1em}
\end{figure}

\subsection{Additional analysis}
To analyze the contributions of individual parts, we conduct ablation experiments with results shown in Fig. \ref{Fig66}.
The D-DFF model achieves 21.53$\%$ encoding time reduction with negligible BDBR loss.
Moreover, the encoding time reduction performances of the D-DFF model in HD and UHD sequences ({\it e.g.} Class A1, Class A2, Class B) are
more significant than that in lower resolution videos.
The encoding time reduction of the P-PBE model remains relatively stable in all sequences with 0.95$\%$ BDBR. Furthermore, the overall algorithm, which incorporates the D-DFF and the P-PBE models, achieves 55.59$\%$ complexity reduction with a tolerable RD loss of 1.40\% on average.

\begin{figure}[h]
	\centering
	\includegraphics[width=0.45\textwidth]{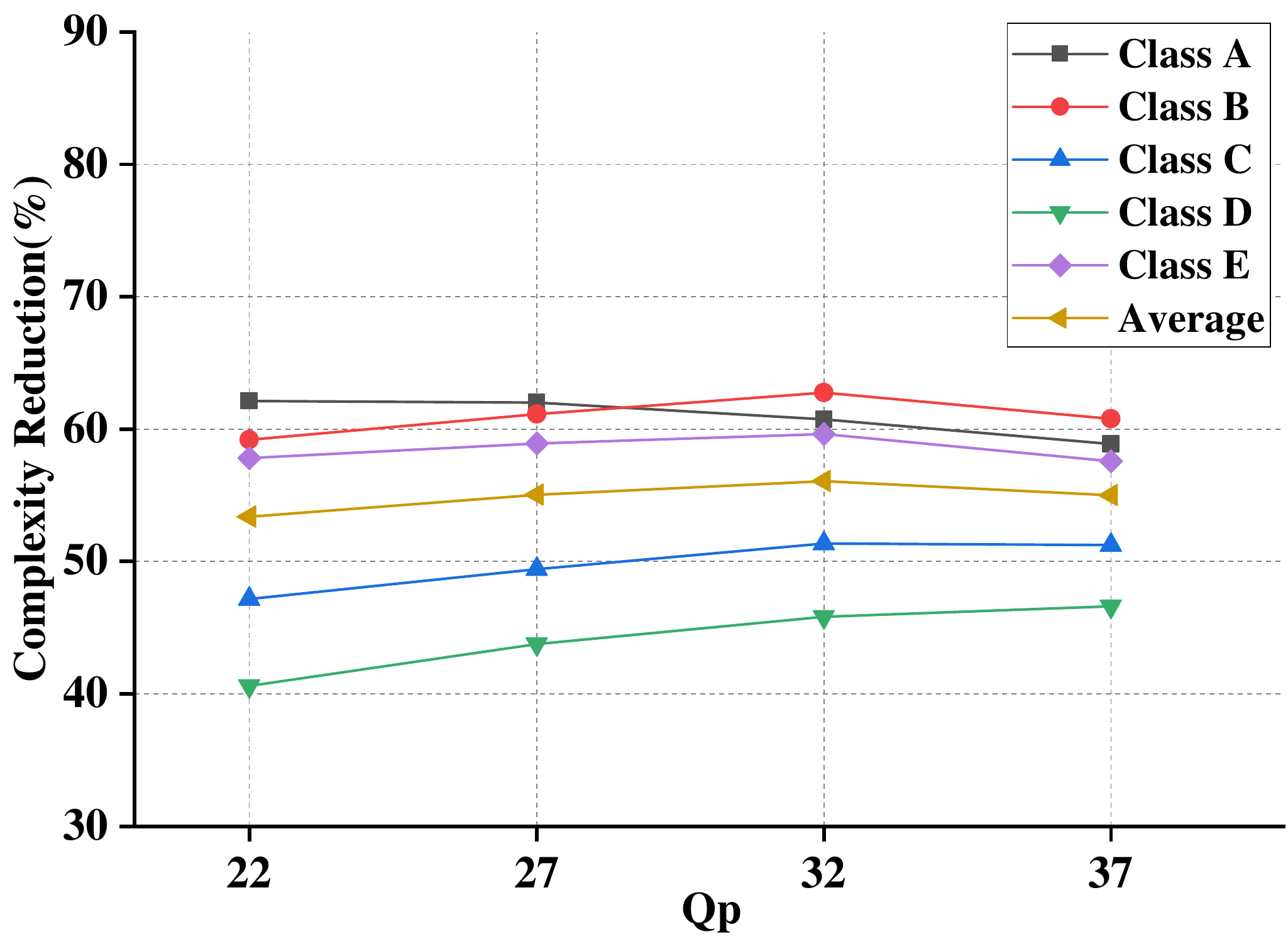}
	\caption{The complexity reduction of different classes under different Qps.\label{Figv42}}
\end{figure}

To analyze the sensitivity of the proposed algorithm with respect to Qp values, we test the time reductions of our algorithm under different Qps and show them in Fig. \ref{Figv42}. It can be observed that the proposed algorithm achieves consistent time saving over different Qps. Therefore, our method is more robust to the changes of coding parameters.

\begin{figure}[h]
	\centering
	\includegraphics[width=0.45\textwidth]{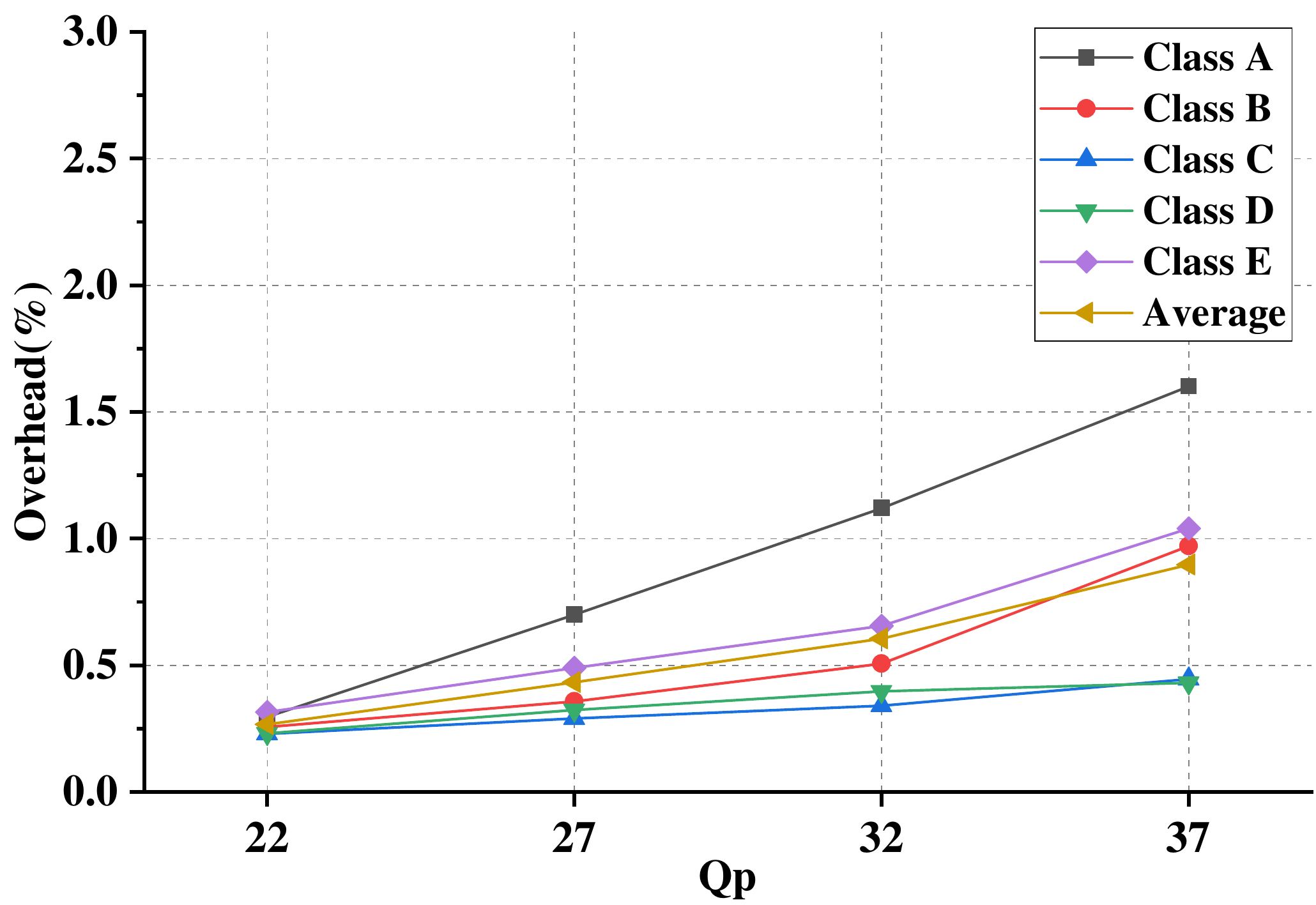}
	\centering
	\caption{ The overhead of proposed algorithm under different classes and Qps.\label{Fig7}}
	\vspace{-1em}
\end{figure}

For completeness, we also analyze the overhead of the proposed algorithm in encoder. 
Fig. \ref{Fig7} shows the overhead of proposed algorithm under different classes and Qps.
Generally, the overhead is generally less than 1.8$\%$, with an average value of 0.48\%. With higher Qp values, the overhead proportion tends to be increased due to the decrease of overall coding complexity. In general, the overhead can be ignored compared with its benefits in overall time reduction.


\section{Conclusion}

This paper addresses the problem of efficient VVC intra coding.
First, we propose a CNN-driven CU depth prediction model and the predicted depth map is used to check early CU termination.
Then, the candidate partition modes are further determined through probability estimation to save more encoding time.
This two-stage approach has a great efficiency of time reduction compared with state-of-the-arts with negligible RD performance loss, offering 55.59$\%$ encoding complexity reduction on average. The whole framework is also applicable in VVC inter prediction with depth-based CU partition. We put it as a future work.

\end{document}